\documentclass[11pt]{article}

\usepackage{color,xspace,wrapfig,fullpage}
\newcommand{\etal}{\emph{et al.}\xspace}
\newcommand{\eg}{\emph{e.g.,}\xspace}
\newcommand{\Eg}{\emph{E.g.,}\xspace}

\usepackage{multirow,booktabs,amsmath,amsfonts,cite,hyperref,float,epsfig,graphics,stfloats,url,bm,booktabs,verbatim,pifont}
\usepackage[hyphenbreaks]{breakurl}
\usepackage{colortbl}
\usepackage{xcolor}
\usepackage{diagbox}

\begin{document}

\title{\LARGE \bf
Contextualize differential privacy in image database: a lightweight image differential privacy approach based on principal component analysis inverse}
	
\author{Shiliang~Zhang\footnote{Chalmers University of Technology, Sweden. Email: shiliang@chalmers.se} \and Xuehui~Ma\footnote{Corresponding author, Xi’an University of Technology, China. Email: maxuehuiphd@stu.xaut.edu.cn} \and Hui Cao\footnote{Xi'an Jiaotong University, China. Email: huicao@mail.xjtu.edu.cn, tyzhao@xjtu.edu.cn, sxczwzz@stu.xjtu.edu.cn} \and Tengyuan Zhao$^\ddag$ \and Yajie Yu\footnote{Henan University of Science and Technology, China. Email: yajieyuw@gmail.com} \and Zhuzhu Wang$^\ddag$ }
	
\maketitle

\newcommand{\tabincell}[2]{\begin{tabular}{@{}#1@{}}#2\end{tabular}}

\begin{abstract}
	
Differential privacy (DP) has been the de-facto standard to preserve privacy-sensitive information in databases. Nevertheless, there lacks a clear and convincing contextualization of DP in the image database, where individual images' indistinguishable contribution to a specific analysis can be achieved and observed when DP is exerted. As a result, the privacy-accuracy trade-off due to integrating DP is insufficiently demonstrated in the context of differentially-private image databases. This work aims at contextualizing DP in an image database by an explicit and intuitive demonstration of integrating conceptional differential privacy with images. To this end, we design a lightweight approach dedicated to privatizing the image database as a whole and preserving the statistical semantics of the image database to an adjustable level, while making individual images' contribution to such statistics indistinguishable. The designed approach leverages principal component analysis (PCA) to reduce the raw image with a large number of attributes to a lower-dimensional space, whereby DP is performed to decrease the DP load of calculating sensitivity attribute-by-attribute. The DP-exerted image data, which is not visible in its privatized format, is visualized through PCA inverse such that both a human and machine inspector can evaluate the privatization and quantify the privacy-accuracy trade-off in an analysis on the privatized image database. Using the devised approach, we demonstrate the contextualization of DP in images by two use cases based on deep learning models, where we show the indistinguishability of individual images induced by DP and the privatized images' retention of statistical semantics in deep learning tasks, which is elaborated by quantitative analyses on the privacy-accuracy trade-off under different privatization settings.

\end{abstract}

\section{Introduction}

We live in a world of data and information, where image data plays an essential role in information sharing and transmitting. Learning from images empowers enormous AI tasks like objective detection, recognition, classification, and reconstruction~\cite{DBLP:conf/csia/YanW21,DBLP:journals/jip/OkumuraHHTK20,DBLP:journals/tip/LiuZBZZ22,DBLP:conf/icassp/YamanHMA21}, whose applications have already permeated the consumer market. However, there has been a growing privacy concern over the recent years as image databases widely accumulated, and machine/deep learning agents strengthen their capacity in extracting information from images, deteriorating the concern further~\cite{DBLP:journals/corr/abs-2103-05472}. The leakage of privacy-sensitive information can induce risks such as unintended human identity revealing~\cite{DBLP:journals/snam/RanaldiZ20,DBLP:conf/sp/HasanCFK20}, personal location/trace exposure~\cite{9633176}, and social/working information disclosure~\cite{DBLP:journals/tifs/YuZKLF17}, raising the crucial need for implementing privacy-preserving approaches.

Among varieties of techniques to protect privacy,~\eg anonymization~\cite{DBLP:journals/ijnc/NakamuraSN21}, encryption~\cite{DBLP:journals/isci/WangY21}, data access control~\cite{DBLP:conf/inista/KountchevMK15}, data outsourcing~\cite{DBLP:journals/isci/YangMMLWM19}, digital forgetting~\cite{DBLP:conf/codaspy/AmjadMP18}, and data summarization~\cite{DBLP:conf/nips/SarpatwarSGJV19}, differential privacy (DP) offers a promising approach to make the contribution of individual data items hardly distinguishable toward given data analyzing tasks~\cite{ziller2021medical}. Such a feature leads to a provable privacy guarantee with a quantitative privacy measurement called privacy budget~\cite{DBLP:journals/pacmpl/BartheCKS021}, making DP the de-facto standard for privacy preservation in data analysis both in academia and industry~\cite{DBLP:journals/tsp/HeCG20,DBLP:conf/mobihoc/YangZHLZ18}. However, though DP has observed its prosperity in a wide range of applications~\cite{DBLP:journals/access/HusnooACDR21,DBLP:journals/corr/abs-2111-02011,DBLP:journals/comsur/HassanRC20,DBLP:journals/tjs/ZhaoZWLU20,DBLP:journals/corr/abs-2010-02973}, its integration into image data is understood in a limited way. Particularly, in the context of images, differential privacy has not gained the conceptional privacy where individual contribution is indistinguishable in a database. Consequently, the privacy-accuracy trade-off due to the adoption of differential privacy in image-data-related analysis has not been fully investigated.

There are efforts to join differential privacy (DP) with images in addressing the mentioned concerns.~\Eg Liu~\etal proposed a facial image differential privacy approach that reconstructs the original image with different components~\cite{DBLP:journals/corr/abs-2103-07073}, \eg mouth or eye, while the use of those components complies with differential privacy. Though this approach guarantees that the outcome image will still be human faces, images processed in that same database are not essentially differentially private from each other. Fan implemented image differential privacy in the granularity of pixels~\cite{DBLP:conf/dbsec/Fan18, Fan2019DifferentialPF}; however, perturbing pixels in individual images does not necessarily lead to quantitative privacy amongst image databases. Fan also performs differential privacy on the singular value decomposition (SVD) of images~\cite{DBLP:conf/icmcs/Fan19}, yet the resulting private images are of low quality and hardly applicable in practice. Quite a few studies leveraged the generative adversary networks (GANs)~\cite{DBLP:journals/jcst/WuYXL19,DBLP:conf/cvpr/SunMOGSF18,DBLP:conf/eccv/SunTXFTS18,DBLP:journals/corr/abs-1912-13457}, in particular, DP-GANs~\cite{DBLP:conf/icvgip/BhattacharjeeBD18a,DBLP:journals/tifs/XuRZZQR19,DBLP:journals/sensors/YuXLWZD21}, to produce a substitute for the original facial image to preserve privacy, and some of those works demonstrated good retention of image semantics from the substitute ones~\cite{DBLP:journals/tip/HeZKSC19,DBLP:conf/iccv/QianLWLWS0H19,DBLP:conf/iccv/LinWCCL19}. GANs also leads to the rise of facial image forgery methods like face synthesis~\cite{DBLP:journals/pami/KarrasLA21}, identity swap~\cite{DBLP:journals/corr/abs-1905-11805}, attribute manipulation~\cite{DBLP:journals/tifs/Gonzalez-SosaFV18}, and expression swap~\cite{DBLP:journals/tog/ThiesZN19}, which can be used to replace sensitive facial semantics to mitigate identity disclosure risk. Nevertheless, the GANs-based approaches mainly contribute to restricting de-identification rather than generating differentially private image databases to be used in, \eg machine learning tasks. Abdur~\etal privatized handwriting images using local differential privacy~\cite{9598983}, which is performed on individual images instead of generating images indistinguishable from each other. Wu~\etal trained a convolutional neural network (CNN) that protects privacy-sensitive information in the training data of pathological images by injecting carefully calculated noise into the CNN model updates~\cite{DBLP:conf/cvpr/WuZSZSZL19}, with their primary focus on the learning of a privacy-aware model. Similarly, Ma~\cite{DBLP:journals/iotj/MaLLMR19} designed a lightweight privacy-preserving mechanism between the images for training and the learning engine, aiming to enhance the learning efficiency while avoiding exposing raw images to the learning engine. Abadi~\etal implemented differential private stochastic gradient descent (DP-SGD)~\cite{DBLP:journals/corr/abs-2106-12576} in the training of a CNN model for image classification~\cite{DBLP:conf/ccs/AbadiCGMMT016}, with the model input dimension reduced by principal component analysis (PCA)~\cite{DBLP:journals/jksucis/BerrimiBH21}. We also note that few works offer intuitive analysis of to what level the differentially privatized images themselves can still be useful in learning tasks like classification, i,e., how privacy trades off accuracy. Chen~\etal studied the privacy-utility trade-off via leveraging differential privacy on facial images~\cite{DBLP:conf/cvpr/ChenCYL21}, yet their approach privatizes images by changing the identifiers of a facial image without demonstrating the indistinguishability between individual image contributions. Guo~\etal implemented differential privacy in decentralized image classification tasks. They analyzed the privacy-accuracy trade-off for the classification performance, while the privacy measures are performed on the learning gradient that feeds the model training rather than on the images themselves~\cite{9515999}. Bernau~\etal conducted differential privacy directly on images and demonstrated the entailed privacy-utility trade-off; however, the privacy measure is carried out on individual images rather than database level~\cite{DBLP:conf/dbsec/BernauRGSK21}.

In this paper, we focus on the contextualization of differential privacy in images regarding individual images' indistinguishable contribution to analyses on an image database and demonstrate an intuitive, explicit, and adjustable privacy-accuracy trade-off for differential privacy on image data. Particularly, we devise a lightweight approach to apply differential privacy on image data, which uses principal component analysis (PCA) to map the raw image into reduced dimensional space whereby we conduct differential privacy and recover the privatized content into its raw dimension to obtain visible differentially-private images. Our approach aims to privatize images in a database to make individual image contributions indistinguishable to both human and machine inspectors. To this end, we test our approach on two learning tasks over images using convolutional neural networks (CNN): handwriting digits recognition and gender classification. Through the experimental results, we depict to what level individual images get indistinguishable from each other as the privacy settings vary and how accurate the trained CNN model can be when differential privacy is exerted on the image database, leading to a detailed insight into the trade-off between the privacy acquirement and the accuracy loss due to the leverage of differential privacy.

This paper is organized as follows: Section~\ref{section: algorithm description} describes the proposed approach in detail. Two CNN experiments for images to test our approach and the corresponding results are shown in section~\ref{section: experiments}. Finally, conclusions are given in section~\ref{section: conclustion}.

\section{Methodology: lightweight image differential privacy using inverse PCA}\label{section: algorithm description}

\begin{figure}[!htbp]
	  \centering
	  \includegraphics[width=0.75\textwidth]{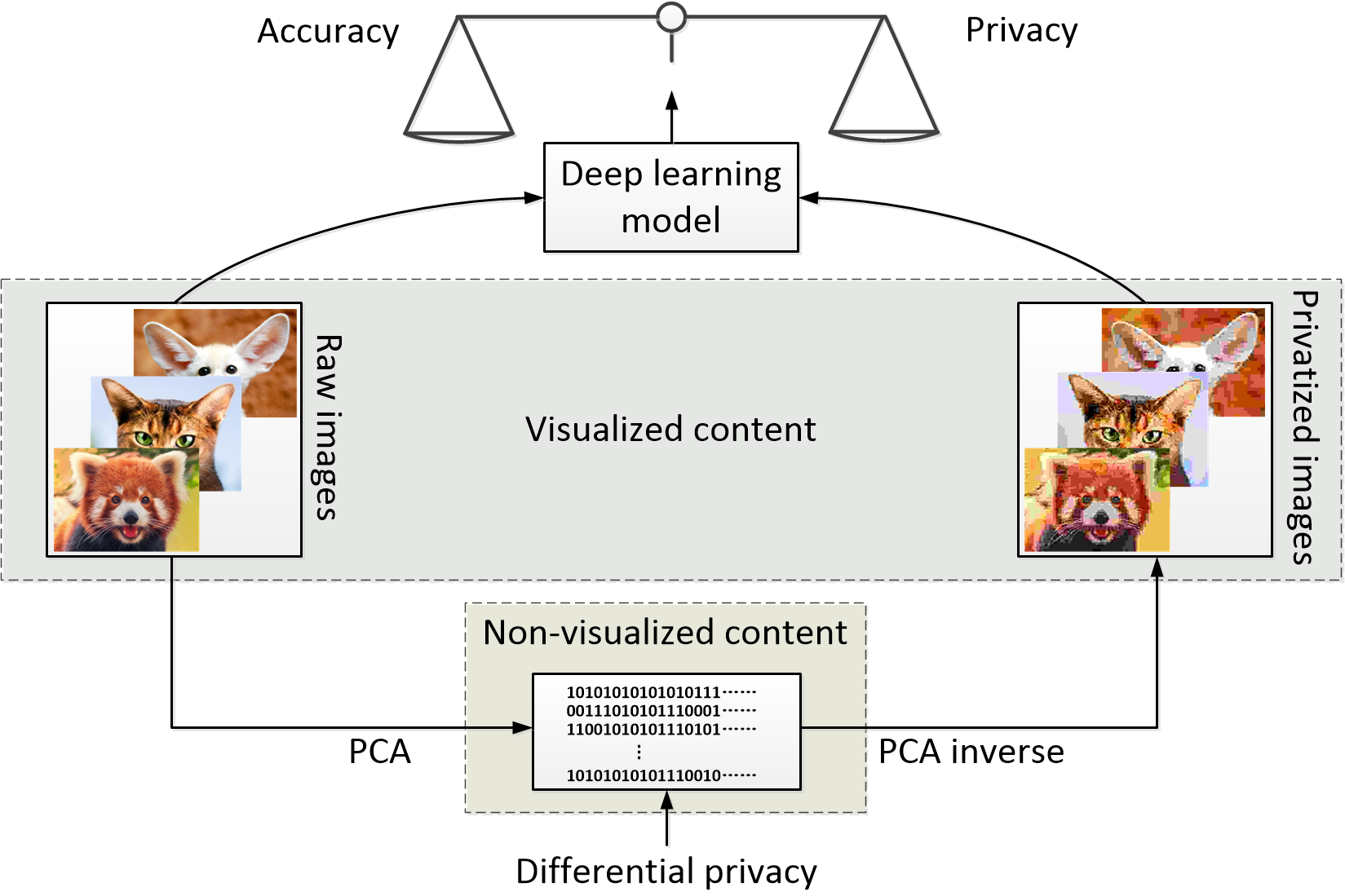}\\
	  \caption{\label{fig: approach overview}Structure of the designed approach}
\end{figure}

Figure~\ref{fig: approach overview} shows a high-level overview of our approach, where we privatize an image database using differential privacy and visualize the privatized images to look into whether their contributions to image-related analyses can be indistinguishable to human inspectors. We also compare the vanilla and the privatized images in deep learning tasks to offer quantitative insights into the trade-off between image privacy and deep-learning-model accuracy due to integrating differential privacy. Our approach leverages PCA as a prerequisite for integrating privacy, the motivation of which is twofold as (i) to reduce image data dimension that leads to less computation for the following differential privacy measures, which calculates the value of sensitivity for every attribute of the database, and (ii) avoid extraneous privatization on irrelevant image content like pure black (or another color) background of a facial image. I.e., our approach conducts differential privacy on the image content that distinguishes individual images from each other most and ignores the image content that means little to the semantics of an image. We detail the design of our method in the following subsections.

\subsection*{a. PCA on image database}

This section illustrates procedures of PCA on an image database to reduce the dimension of images. We conduct PCA on images as a first and preparing step for image privatization. Suppose we have an image database $D=\{I_{1}, I_{2}, ... I_{n}\}$, where $I_{i}, i\in\{1,2,...,n\}$ is an image with $m\times m$ pixels. By reconstructing each image in this database into a vector, we can obtain the image database as $D'=\{I'_{1}, I'_{2}, ... I'_{n}\}$, where

\begin{equation}\label{equation: image flat}
 	\begin{split}
 		I'_{i}=\left[p^{i}_{1},\; p^{i}_{2},\; ...\;,\; p^{i}_{s}\right]^{T},\; i\in\{1, 2, ... , n\},\; and\; s=m^{2}\\
 	\end{split}
 \end{equation}

Then we mean-centralize the reconstructed images by subtracting the mean vector for the images by

\begin{equation}\label{equation: mean centralized}
	\begin{split}
		\widehat{I}'_{i}=\left[p^{i}_{1},\; p^{i}_{2},\; ...\;,\; p^{i}_{s}\right]^{T}-\left[\bar{p}_{1},\; \bar{p}_{2},\; ...\;,\; \bar{p}_{s}\right]^{T}\\
	\end{split}
\end{equation}
where

\begin{equation}\label{equation: mean}
	\begin{split}
		\bar{p}_{j}=\frac{1}{n}\sum_{k=1}^{n}p^{k}_{j},\; j\in\{1, 2, ... , s\}\\
	\end{split}
\end{equation}

After that, the data matrix of mean-centralized images $\widehat{D}'=\{\widehat{I}'_{1}, \widehat{I}'_{2}, ... \widehat{I}'_{n}\}$ is multiplied by its transpose to calculate the covariance matrix as

\begin{equation}\label{equation: covariance matrix}
	\begin{split}
		C=\widehat{D}'\widehat{D}'^{T}\\
	\end{split}
\end{equation}

Assume the eigenvalue matrix for the covariance matrix $C$ is

\begin{equation}\label{equation: eigenvalue}
\Lambda=\left[\begin{split}
\lambda_{1} \quad 0 \quad ...  \quad 0\\
0 \quad \lambda_{2} \quad ...  \quad 0\\
...\quad \quad \quad \\
0 \quad 0 \quad ...  \quad \lambda_{s}
\end{split}
\right]\\
\end{equation}
where $\lambda_{1}\geq\lambda_{2}\geq ...\geq\lambda_{s}$, and the corresponding eigenvector matrix is $E=\left[e_{1},e_{2},...,e_{i},...,e_{s}\right]$ where $e_{i}$ is a ($s\times 1$)-dimension eigenvector corresponding to $\lambda_{i}$. Given the objective dimension $d\; (d<s)$ to which we want to reduce the raw image, the dimension-reduced image database can be obtained as $D^{r}=\{I^{r}_{1}, I^{r}_{2}, ... I^{r}_{n}\}$, where

\begin{equation}\label{equation: reduced image}
	\begin{split}
		I^{r}_{i}=(\widehat{I}'_{i})^{T}E^{r}=\left[\widetilde{p^{i}_{1}},\widetilde{p^{i}_{2}},...,\widetilde{p^{i}_{d}}\right], i\in\{1,2,...,n\}\\
	\end{split}
\end{equation}
\begin{equation}\label{equation: eigenvector}
	\begin{split}
		E^{r}=\left[e_{1},e_{2},...,e_{d}\right]\\
	\end{split}
\end{equation}

Consequently, each image is now reduced to a vector with $d$ attributes and no longer visualized, and those reduced attributes represent the information that distinguishes individual image data from each other most. Therefore, when exerting differential privacy on the reduced image data, our efforts will be focused on privatizing the image content that significantly contributes to the semantics of an image.

\subsection*{b. Differential privacy for the dimension-reduced image data}

This subsection describes the crucial step toward a privatized image database where individual image data contributes indistinguishably to an analysis. For a given database $\mathbb{D}$, differential privacy sets a privacy budget $\epsilon$, depending on which a sequence of noise is quantified and injected into each item in the database. Thereafter, for two databases $DB\subseteq\mathbb{D}$ and $DB'\subseteq\mathbb{D}$ that differ from each other by only one data entry, the data analysis $A$ is said to give $\epsilon$-differential privacy for all events $S\subseteq output(A)$, such that

\begin{equation}\label{equation: differential privacy}
	\begin{split}
		Pr\left[A(DB)\subseteq S\right]\leq e^{\epsilon}\cdot Pr\left[A(DB')\subseteq S\right]\\
	\end{split}
\end{equation}
where $output(\;)$ denotes the range of output with a given input, $Pr[\;]$ means probability distribution. In this way, individual data items' contribution to analysis $A$ is hardly distinguishable; thus, their privacy in a database can be preserved.

Our approach applies differential privacy to the dimension-reduced image data to pursue a privatized image database. The key procedure here is to calculate the right amount of noise according to the privacy budget $\epsilon$ and inject the noise into the data, which determines the balance of the privacy-accuracy trade-off in the subsequent analysis upon the data. This work uses Laplace noise, with its mean value as zero and the variance calculated separately for each attribute in the dimension-reduced image database $D^{r}$ obtained through equation~\ref{equation: reduced image}. Specifically, for the noise $n_{l}\; (l\in\{1,2,...,d\})$ to be injected into the $l$-th attribute, the Laplace noise scale  $\lambda^{l}$ is calculated by 

\begin{equation}\label{equation: noise calculation}
	\begin{split}
		\lambda^{l}=\frac{sensitivity(l)}{\epsilon},\;l\in\{1,2,...,d\}\\
	\end{split}
\end{equation}
\begin{equation}\label{equation: sensitivity calculation}
	\begin{split}
		sensitivity(l)=||max(\widetilde{p^{i}_{l}})-min(\widetilde{p^{j}_{l}})||_{1},\;i\;and\; j\in\{1,2,...,n\},\;i\neq j\\
	\end{split}
\end{equation}
where $\widetilde{p^{i}_{l}}$ and $\widetilde{p^{j}_{l}}$ mean the value for the $l$-th attribute in the $i$-th and $j$-th reduced image data, respectively, and $sensitivity(l)$ describes the range that the value of the $l$-th attribute can differ in the database. With the distribution of the noise $n_{l}$ settled, noise can be generated and injected to obtain the differentially private image database $D^{r}_{dp}=\{I^{dp}_{1},I^{dp}_{2},...,I^{dp}_{n}\}$, where

\begin{equation}\label{equation: differentially private database}
	\begin{split}
		I^{dp}_{i}=\left[\widetilde{p^{i}_{1}},\widetilde{p^{i}_{2}},...,\widetilde{p^{i}_{d}}\right]+\left[rand(n_{1}),rand(n_{2}),...,rand(n_{d})\right], i\in\{1,2,...,n\}\\
	\end{split}
\end{equation}
in which $rand(\;)$ represents a random value generated from the distribution of a given noise.

Now we have gained the privatized image database that complies with $\epsilon$-differential privacy. Nevertheless, the obtained database is not visualized, i.e., the processed images are no longer in an image format; instead, they are privatized vectors encoding the principle components of image semantics. Therefore, there is a need to convert the privatized images to a visible format such that both human inspectors and machine inspectors can look into differential privacy's impact on the images and further analyze the privacy-accuracy trade-off. The visualization of the privatized images is illustrated in the following subsection.

\subsection*{c. PCA inverse for visualization of the privatized images}

Upon the previous steps, we obtain privatized images with a reduced dimension of $1\times d$, and we will recover the images to their raw dimensional space $n\times n$ to make them visible. 

The initial step here is to map the image data from $1\times d$ to $1\times s$ dimensional space to retrieve all the attributes lost because of PCA, which we call PCA inverse. As illustrated in equation~\ref{equation: reduced image}, the image data dimension is reduced through $I^{r}_{i}=(\widehat{I}'_{i})^{T}E^{r}$, based which we can derive the inverse of this reduction as:

\begin{equation}\label{equation: pca inverse}
	\begin{split}
		I^{r}_{i}(E^{r})^{T}&=(\widehat{I}'_{i})^{T}E^{r}(E^{r})^{T} \\
		I^{r}_{i}(E^{r})^{T}\left(E^{r}(E^{r})^{T}\right)^{-1}&=(\widehat{I}'_{i})^{T}E^{r}(E^{r})^{T}\left(E^{r}(E^{r})^{T}\right)^{-1} \\
		(\widehat{I}'_{i})^{T}&=I^{r}_{i}(E^{r})^{T}\left(E^{r}(E^{r})^{T}\right)^{-1}
	\end{split}
\end{equation}
To circumvent a null solution, we add a regulator matrix $\Lambda_{inv}$ consisting of tiny positive values in the matrix inverse operation that the derivation in equation~\ref{equation: pca inverse} becomes:

\begin{equation}\label{equation: regularized pca inverse}
	\begin{split}
		(\widehat{I}'_{i})^{T}&=I^{r}_{i}(E^{r})^{T}\left(E^{r}(E^{r})^{T}+\Lambda_{inv}\right)^{-1}\\
		\Lambda_{inv}&=\left[\begin{split}\lambda^{inv}_{1} \quad 0 \quad ...  \quad 0\\
		0 \quad \lambda^{inv}_{2} \quad ...  \quad 0\\
		...\quad \quad \quad \\
		0 \quad 0 \quad ...  \quad \lambda^{inv}_{s}
	\end{split}\right]
	\end{split}
\end{equation}
Since $\widehat{I}'_{i}$ is mean-centralized as shown in equation~\ref{equation: mean centralized}, the mean vector $\left[\bar{p}_{1},\; \bar{p}_{2},\; ...\;,\; \bar{p}_{s}\right]^{T}$ need to be added to obtain $I'_{i}$ as

\begin{equation}\label{equation: de-mean-centralization}
	\begin{split}
		(I'_{i})^{T}&=(\widehat{I}'_{i})^{T}+\left[\bar{p}_{1},\; \bar{p}_{2},\; ...\;,\; \bar{p}_{s}\right]\\
		&=I^{r}_{i}(E^{r})^{T}\left(E^{r}(E^{r})^{T}+\Lambda_{inv}\right)^{-1}+\left[\bar{p}_{1},\; \bar{p}_{2},\; ...\;,\; \bar{p}_{s}\right]\\
		I'_{i}&=\left(I^{r}_{i}(E^{r})^{T}\left(E^{r}(E^{r})^{T}+\Lambda_{inv}\right)^{-1}\right)^{T}+\left[\bar{p}_{1},\; \bar{p}_{2},\; ...\;,\; \bar{p}_{s}\right]^{T}
	\end{split}
\end{equation}
Then reshape the $(s\times 1)$-dimension vector $I'_{i}$ to an $n\times n$ matrix, and we manage to recover the dimension-reduced image data to its raw visible format. Note that by replacing $I^{r}_{i}$ with the reduced \& privatized image data $I^{dp}_{i}$ in equation~\ref{equation: de-mean-centralization}, we can obtain the privatized and visualized version of images that can be either scrutinized by a human inspector or be used to train/test a machine learning or deep learning model.

\section{Implementation and experimental results}\label{section: experiments}

This section verifies our designed approach through two demonstrations. The first one trains a CNN model to classify handwriting digits using the MNIST database~\footnote{\url{http://yann.lecun.com/exdb/mnist/}}, and the second one trains a more sophisticated CNN model to recognize human gender from facial images, using an open-source facial database from GitHub~\footnote{\url{https://github.com/esengie/GenderRecog}}. Table~\ref{table: data descriptions} describes the two data sets and their specifications in the model training. With the two databases, we train both models in the two experiments in a non-private way while we test the trained model using the images privatized by our approach. Thus, in the testing phase, we can look into what level the CNN model accuracy decreases for specified privacy settings and depict a clear privacy-accuracy trade-off for differential privacy on images. We conduct all the model training and testing using MATLAB R2021b with access to NVidia Tesla T4 GPUs.

\begin{table}[htp]
	\centering
	\caption{\label{table: data descriptions}Data specifications for the experiments.}
	\resizebox{0.9\textwidth}{!}{
\begin{tabular}{|l|c|c|c|c|c|}
			\cline{2-6}
			 \multicolumn{1}{c|}{} & \multirow{2}{*}{\tabincell{c}{Image\\type}}  & \multirow{2}{*}{\tabincell{c}{Image\\resolution}} & \multirow{2}{*}{\tabincell{c}{Labels}} & \multicolumn{2}{c|}{Database size} \\
			 \cline{5-6}
			 \multicolumn{1}{c|}{} & & & & For taining & For testing \\
			\hline
			 Handwriting digit database & Grayscale & 28$\times$ 28 pixels & 0-9 integers & 60,000 & 10,000\\
			 \hline
			 Facial database & 24-bit color & 250$\times$ 250 pixels & Female or male & 4,709 & 1,177 \\
			\hline
		\end{tabular}
	}
\end{table}

\subsection{Experiment 1: handwriting digits recognition}

\begin{table}[h!]
	\centering
	\resizebox{0.99\textwidth}{!}{
		\setlength\arrayrulewidth{1.2pt}
		\begin{tabular}{ !{\color{red}\vrule width 1.2pt} c!{\color{red}\vrule  width 1.2pt} c!{\color{red}\vrule  width 1.2pt} c!{\color{red}\vrule  width 1.2pt} c!{\color{red}\vrule  width 1.2pt} c!{\color{red}\vrule  width 1.2pt} c!{\color{red}\vrule  width 1.2pt} c!{\color{red}\vrule  width 1.2pt} c!{\color{red}\vrule  width 1.2pt} c!{\color{red}\vrule  width 1.2pt} c!{\color{red}\vrule  width 1.2pt} c!{\color{red}\vrule  width 1.2pt} c!{\color{red}\vrule  width 1.2pt} c!{\color{red}\vrule  width 1.2pt} c!{\color{red}\vrule  width 1.2pt} c!{\color{red}\vrule  width 1.2pt} c!{\color{red}\vrule  width 1.2pt} c!{\color{red}\vrule  width 1.2pt} c!{\color{red}\vrule  width 1.2pt} c!{\color{red}\vrule  width 1.2pt} c!{\color{red}\vrule  width 1.2pt} c!{\color{red}\vrule  width 1.2pt}    }
			\arrayrulecolor{red}\hline
			\diagbox[innerwidth=5mm]{$d$}{$\epsilon$} & $1$ &$2$  &$3$  &$4$  &$5$  & $6$ &$7$  &$8$  & $9$ &$10$  & $11$ & $12$ & $13$ &$14$  & $15$ &$16$  &$17$  & $18$ & $19$ & $20$ \\
\hline
		\end{tabular}
	}
	\caption{Privatized images for the handwriting digit labeled as ``2" in the MNIST database. The table cells highlighted in green are examples that look similar to its non-privatized version from human's vision.}\label{table: dp individual image}
\end{table}

\subsubsection{Model training}
Using the raw training images of the handwriting digit data, we learn a CNN model to classify which integer the digit images show, and the classification accuracy is validated by the raw testing images. The learned CNN model consists of two convolution layers each with five $[5, 5]$ filters, and two pooling layers each with the pooling region dimension of $[2,2]$. In the vanilla training and testing phase, we gain the model accuracy of 99.05\% and 98.68\%, respectively. Then we privatize the images using the proposed approach and test the obtained model using the privatized images. We take the accuracy for the vanilla testing phase as the benchmark to check how the privatized testing images compromise accuracy and privacy.

\begin{table}[h!]
	\centering
	\resizebox{0.99\textwidth}{!}{
		\setlength\arrayrulewidth{1.2pt}
\begin{tabular}{ !{\color{red}\vrule width 1.2pt} c!{\color{red}\vrule  width 1.2pt} c!{\color{red}\vrule  width 1.2pt} c!{\color{red}\vrule  width 1.2pt} c!{\color{red}\vrule  width 1.2pt} c!{\color{red}\vrule  width 1.2pt} c!{\color{red}\vrule  width 1.2pt} c!{\color{red}\vrule  width 1.2pt} c!{\color{red}\vrule  width 1.2pt} c!{\color{red}\vrule  width 1.2pt} c!{\color{red}\vrule  width 1.2pt} c!{\color{red}\vrule  width 1.2pt} c!{\color{red}\vrule  width 1.2pt} c!{\color{red}\vrule  width 1.2pt} c!{\color{red}\vrule  width 1.2pt} c!{\color{red}\vrule  width 1.2pt} c!{\color{red}\vrule  width 1.2pt} c!{\color{red}\vrule  width 1.2pt} c!{\color{red}\vrule  width 1.2pt} c!{\color{red}\vrule  width 1.2pt} c!{\color{red}\vrule  width 1.2pt} c!{\color{red}\vrule  width 1.2pt}    }
			\arrayrulecolor{red}\hline
			Raw &$\epsilon=1$  &$\epsilon=2$  &$\epsilon=3$  &$\epsilon=4$  & $\epsilon=5$ &$\epsilon=6$  &$\epsilon=7$  & $\epsilon=8$ &$\epsilon=9$  & $\epsilon=10$ & $\epsilon=11$ & $\epsilon=12$ &$\epsilon=13$  & $\epsilon=14$ &$\epsilon=15$  &$\epsilon=16$  & $\epsilon=17$ & $\epsilon=18$ & $\epsilon=19$ & $\epsilon=20$ \\
\hline
		\end{tabular}
	}
	\caption{An example of privatized images for the handwriting digit labeled from ``0" to ``9" in the MNIST database given the reduced image dimension $d=20$. The first column shows the non-privatized digit images. The background of table cells highlighted by green are examples that look similar to its non-privatized version from human's vision.}\label{table: dp multi image}
\end{table}

\subsubsection{Image privatization and model testing based on differentially-private images}
\begin{figure}[!htbp]
	\centering
	\includegraphics[width=0.95\textwidth]{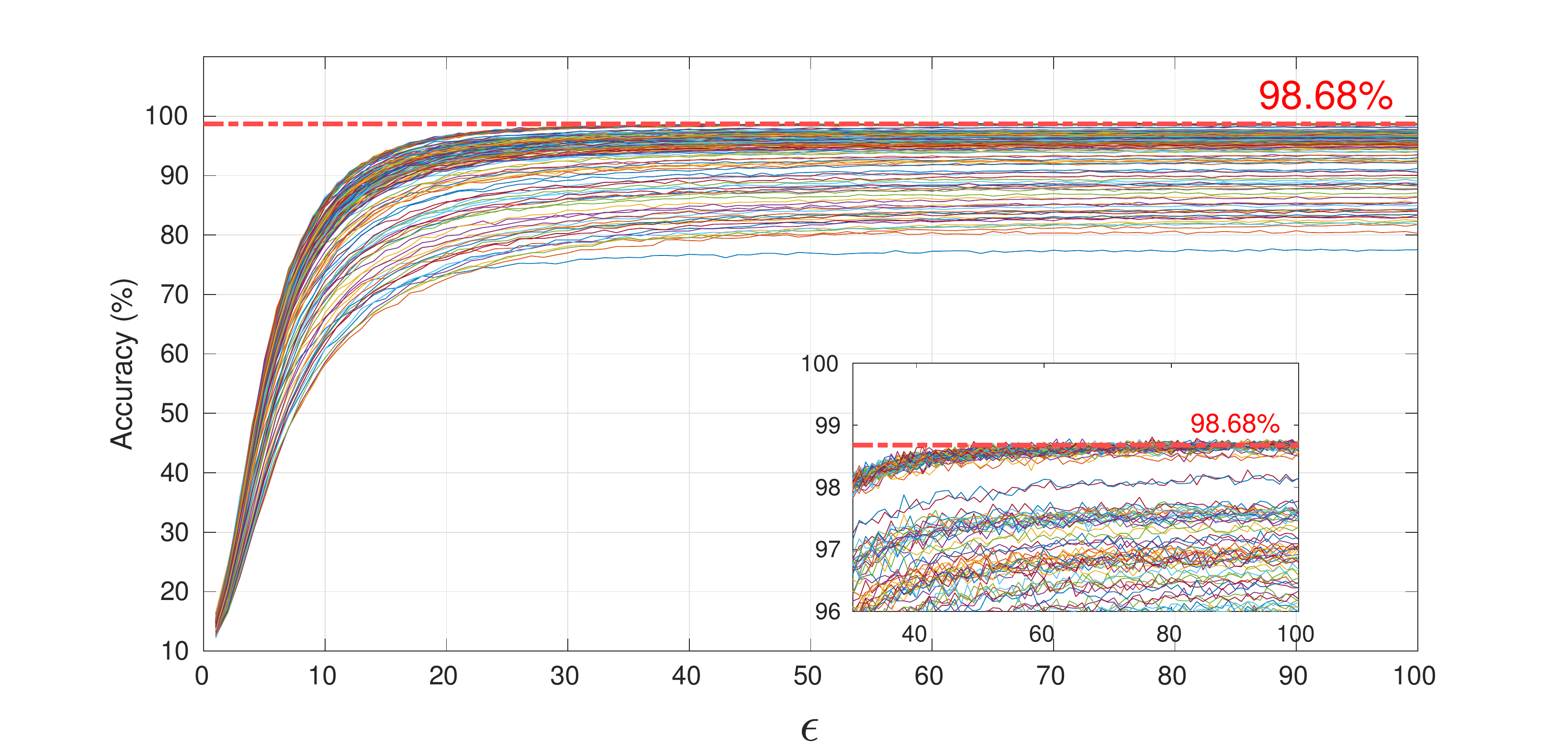}\\
	\caption{\label{fig: accuracy epsilon}Impact of $\epsilon$ on testing accuracy for handwritting digit recognition. Each of the colored curves represents the plot of model test accuracy under a given reduced image dimension $d$, $d\in\{10,11,12,13,...,198,199,200\}$. The horizontal line at $accuracy=98.68\%$ means the reference accuracy obtained by testing the CNN model with non-privatized images.}
\end{figure}

\begin{figure}[!htbp]
	\centering
	\includegraphics[width=0.95\textwidth]{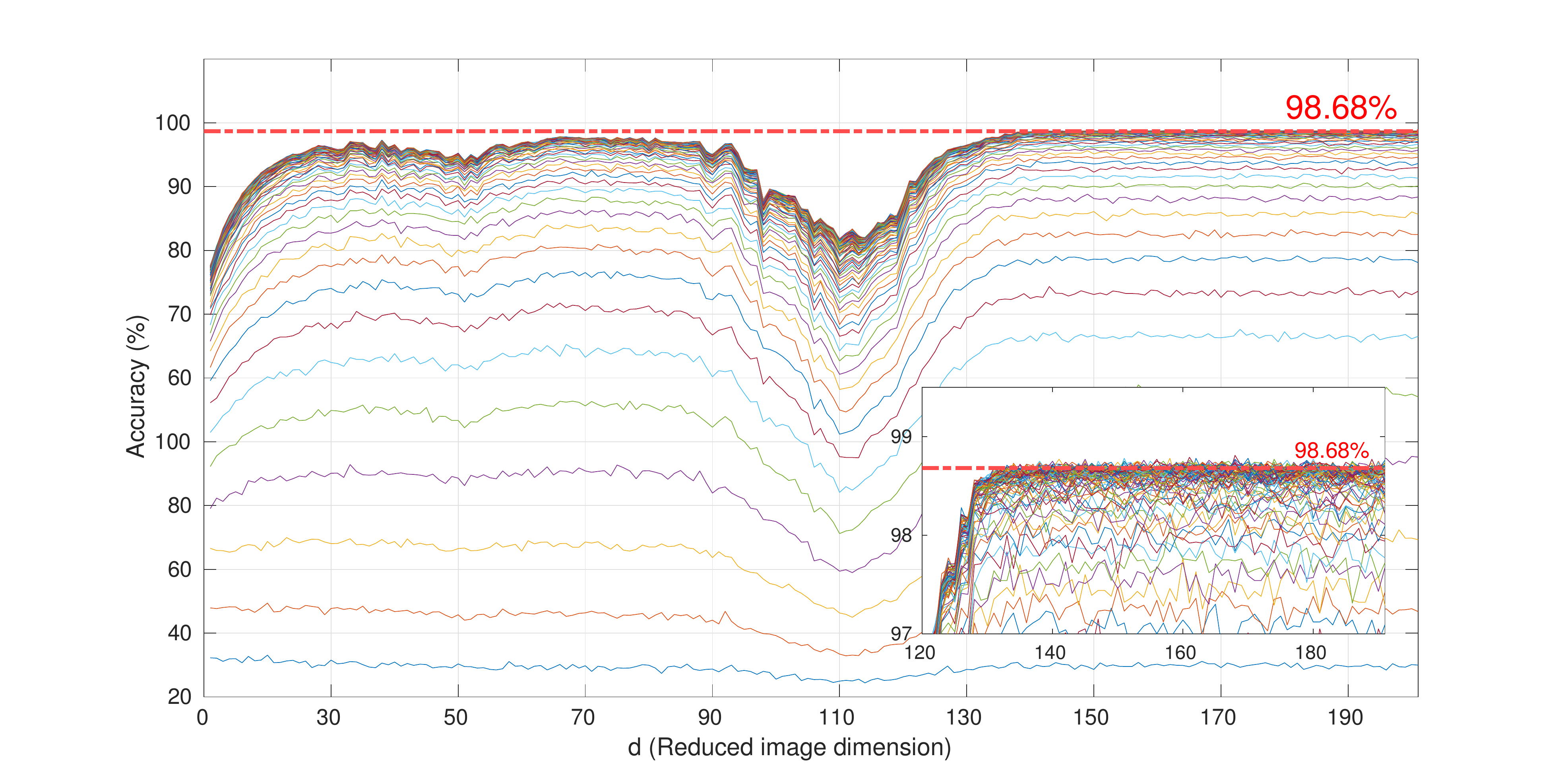}\\
	\caption{\label{fig: accuracy dimension}Impact of reduced image dimension $d$ on testing accuracy for handwriting digit recognition. Each of the colored curves represents the plot of model test accuracy under a given privacy budget $\epsilon$, $\epsilon\in\{1,2,3,4,...,98,99,100\}$. The horizontal line at $accuracy=98.68\%$ means the reference accuracy obtained by testing the CNN model with non-privatized images.}
\end{figure}

This subsection visualizes the differentially-private images under different privatizing settings and demonstrates how the privacy-accuracy trade-off appears over those settings. Specifically, we vary (i) the privacy budget $\epsilon$ and (ii) the reduced number of attributes $d$ (for the PCA operation, as shown in equation~\ref{equation: reduced image}) during the experiment. To accelerate the PCA calculation on image data with a large number of attributes, we privatize the image database in a batch of 100 images, and show how the privatized images are perceived by both human vision and machine learning inspectors. Table~\ref{table: dp individual image} how the appearance of a single image differs over different combinations of $\epsilon$ and $d$, while Table~\ref{table: dp multi image} shows how different images get indistinguishable from a database perspective. Both Table~\ref{table: dp individual image} and~\ref{table: dp multi image} manifest a trend that the level of perturbation on the semantics of the image lowers as privacy budget $\epsilon$ increases, indicating less accuracy loss regarding an image analysis. Using an already trained CNN model to classify the privatized testing images and looking into the classification accuracy, Figures~\ref{fig: accuracy epsilon} and~\ref{fig: accuracy dimension} provide a quantitative analysis of the privacy-accuracy trade-off under different settings of $\epsilon$ and $d$.

It can be observed from figure~\ref{fig: accuracy epsilon} that larger $\epsilon$ brings higher accuracy in the CNN model testing, which is as anticipated since larger $\epsilon$ induces less noise injected into the image to distort the accuracy during the model testing. However, the reduced image dimension $d$ shows a non-monotone impact, and there exists a valley for $d$ around $[80,~120]$ where the accuracy decreases with $d$ at the beginning of the valley and then increase, as shown in Figure~\ref{fig: accuracy dimension}. Such a trend implicates that in this experiment, simply using more attributes of an image does not necessarily benefit the retention of image semantics (or accuracy in this experiment). Instead, a ``pitfall" region seems to exist that the noise-injected images whose dimension is reduced to this region can even confuse the CNN model more and incurs lower accuracy. We also notice from Figures~\ref{fig: accuracy epsilon} and~\ref{fig: accuracy dimension} that the model testing accuracy on the privatized testing images is either below or can not be stably above the reference accuracy obtained from the vanilla images, which manifests the price to pay for the acquisition of privacy.

\subsection{Experiment 2: gender classification based on human facial image}

\subsubsection{Model training}

This experiment trains a CNN model with seven convolution layers each with twenty $[5, 5]$ filters, seven batch normalization layers to normalize the data stream and speed up the training, seven rectified linear unit layers, and seven average pooling layers each with a pooling size of $[2, 2]$. The model training uses a human facial image database, and the trained model can recognize gender for a given input image. As a pre-processing, we reduce images in the database to the size ($80\times 80$) pixels to accelerate the training and testing. The training and testing accuracy for this CNN model is 99.89\% and 90.24\%, respectively, using the vanilla images. Then we test the trained model using privatized images and check the resulting privacy-accuracy trade-off under different settings. During the model testing on privatized images, we use the accuracy from vanilla testing images as the reference to see to what level the accuracy decreases due to integrating differential privacy.

\subsubsection{Image privatization and model testing based on differentially-private images}

We privatize the human facial image database under different combinations of privacy budget $\epsilon$ and the reduced image dimension $d$, and look into how the privatized images are perceived both by human vision and machine learning inspector. First, we show how privatized images appear both from an individual image and database perspective, followed by a quantitative analysis of the privacy-accuracy trade-off by testing the trained CNN model using the privatized testing images. Note that we use a batch of 300 images in the privatization procedures to speed up the PCA calculation on image data with a considerable amount of attributes.

\begin{table}[h!]
	\centering
	\resizebox{0.99\textwidth}{!}{
		\setlength\arrayrulewidth{1.2pt}
\begin{tabular}{ !{\color{red}\vrule width 1.2pt} c!{\color{red}\vrule  width 1.2pt} c!{\color{red}\vrule  width 1.2pt} c!{\color{red}\vrule  width 1.2pt} c!{\color{red}\vrule  width 1.2pt} c!{\color{red}\vrule  width 1.2pt} c!{\color{red}\vrule  width 1.2pt} c!{\color{red}\vrule  width 1.2pt} c!{\color{red}\vrule  width 1.2pt} c!{\color{red}\vrule  width 1.2pt} c!{\color{red}\vrule  width 1.2pt} c!{\color{red}\vrule  width 1.2pt} c!{\color{red}\vrule  width 1.2pt} c!{\color{red}\vrule  width 1.2pt} c!{\color{red}\vrule  width 1.2pt} c!{\color{red}\vrule  width 1.2pt} c!{\color{red}\vrule  width 1.2pt}   }
			\arrayrulecolor{red}\hline
			\diagbox[innerwidth=5mm]{$d$}{$\epsilon$} & $1$ &$2$  &$3$  &$4$  &$5$  & $6$ &$7$  &$8$  & $9$ &$10$  & $11$ & $12$ & $13$ &$14$  & $15$  \\
\hline
		\end{tabular}
	}
	\caption{Privatized version for a single human facial image.}\label{table: dp individual facial image}
\end{table}

\begin{figure}[!htbp]
	\centering
	\includegraphics[width=0.75\textwidth]{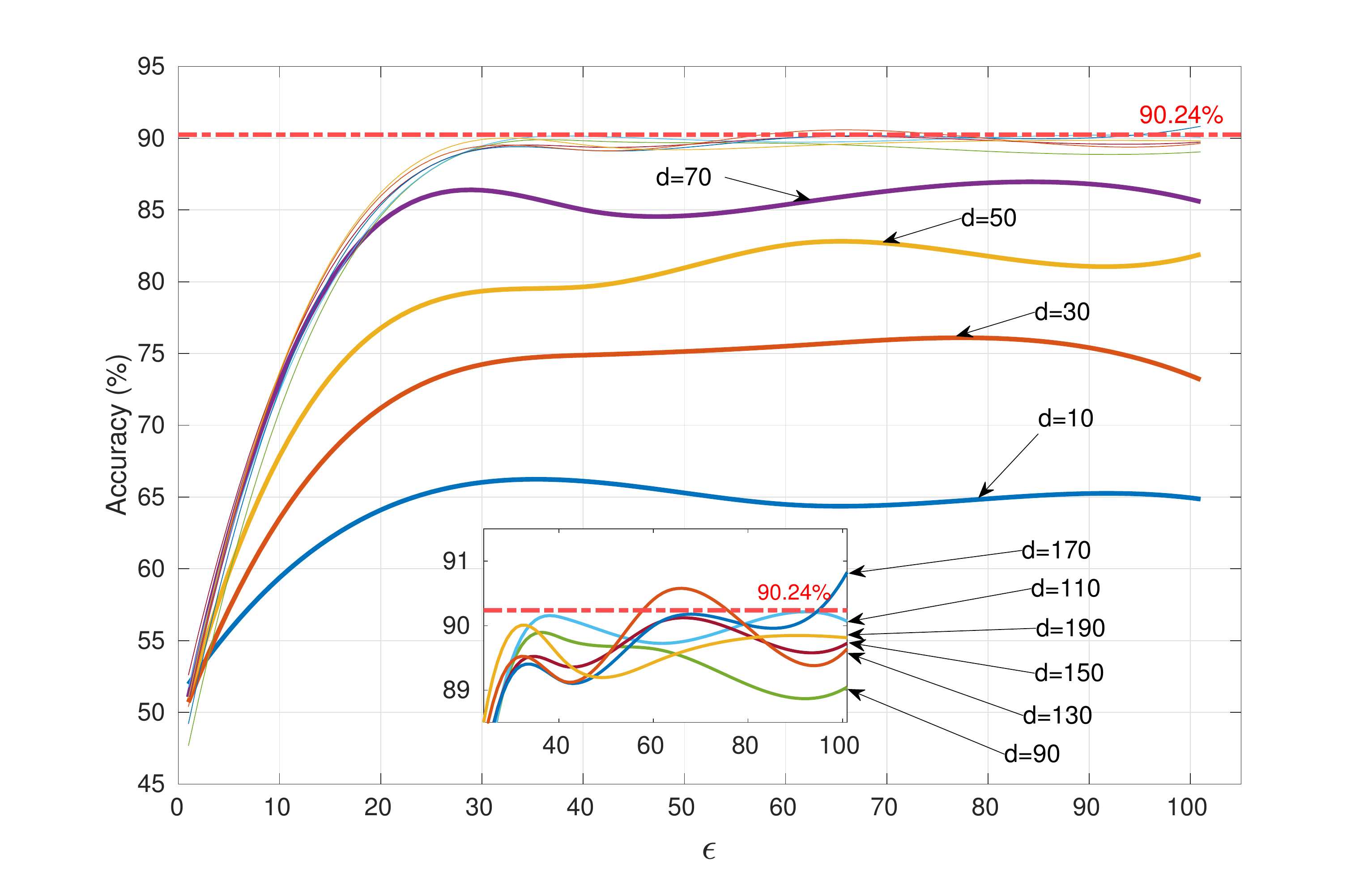}\\
	\caption{\label{fig: dp impact on cnn facial}Model testing accuracy for the CNN gender recognition regarding $\epsilon$ and $d$. The horizontal line at $accuracy=90.24\%$ means the reference accuracy obtained by testing the CNN model with non-privatized images}
\end{figure}

Table~\ref{table: dp individual facial image} presents the appearance of an individual image privatized under different $\epsilon$ and $d$, where $\epsilon$ tends to impact the image semantics more significantly than the reduced image dimension $d$, while higher $d$ tends to bring better image semantics slightly, as shown in column $d=4$ and $d=5$. Table~\ref{table: dp multiple facial image} shows how images from the facial database get more indistinguishable/distinguishable from each other as the privacy budget varies, and it also manifests a clear trend that lower $\epsilon$ makes it harder to tell one from another. Figure~\ref{fig: dp impact on cnn facial} provides a quantitative insight into the privacy-accuracy trade-off via testing the trained CNN model using privatized testing images. Note that the plots in Figure~\ref{fig: dp impact on cnn facial} are smoothed using the MATLAB built-in spline function to avoid oscillations that hardly damage or contribute to the result analysis. 

\begin{table}[h!]
	\centering
	\resizebox{1\textwidth}{!}{
		\setlength\arrayrulewidth{1.2pt}
\begin{tabular}{ !{\color{red}\vrule width 1.2pt} c!{\color{red}\vrule  width 1.2pt} c!{\color{red}\vrule  width 1.2pt} c!{\color{red}\vrule  width 1.2pt} c!{\color{red}\vrule  width 1.2pt} c!{\color{red}\vrule  width 1.2pt} c!{\color{red}\vrule  width 1.2pt} c!{\color{red}\vrule  width 1.2pt} c!{\color{red}\vrule  width 1.2pt} c!{\color{red}\vrule  width 1.2pt} c!{\color{red}\vrule  width 1.2pt} c!{\color{red}\vrule  width 1.2pt} c!{\color{red}\vrule  width 1.2pt} c!{\color{red}\vrule  width 1.2pt} c!{\color{red}\vrule  width 1.2pt} c!{\color{red}\vrule  width 1.2pt} c!{\color{red}\vrule  width 1.2pt}   }
			\arrayrulecolor{red}\hline
			raw & $\epsilon=1$ &$\epsilon=2$  &$\epsilon=3$  &$\epsilon=4$  &$\epsilon=5$  & $\epsilon=6$ &$\epsilon=7$  &$\epsilon=8$  & $\epsilon=9$ &$\epsilon=10$  & $\epsilon=11$ & $\epsilon=12$ & $\epsilon=13$ &$\epsilon=14$  & $\epsilon=15$  \\
\hline
			
		\end{tabular}
	}
	\caption{An example of privatized image database with a database size of 18 regarding different values of $\epsilon$, given a fixed reduced image dimension of $d=20$.}\label{table: dp multiple facial image}
\end{table}

Figure~\ref{fig: dp impact on cnn facial} shows clear patterns that the testing accuracy grows with both $\epsilon$ and $d$, which is quite different from the results on the handwritting digits where a ``pitfall" region exists for the reduced image dimension $d$. Such a difference indicates that for the human facial images under a fixed privacy budget $\epsilon$ and the consequent privatization, using more attributes extracted from PCA contributes to the retention of image semantics from a machine learning perspective. We can also conclude further that whether more attributes mean more semantics differs from database to database, according to the quality and richness of the images.

\section{Conclusions}\label{section: conclustion}

This paper contextualizes the conceptional differential privacy in image databases by designing and implementing a lightweight image differential privacy approach, and a precise analysis on the privacy-accuracy trade-off due to integrating differential privacy. The devised approach is dedicated to privatizing an image database as a whole and making individual image contributions indistinguishable, rather than privatizing individual images separately. Thus, we can hold the statistical semantics of an image database to a certain level adjusted by the privatization settings, which gets consistent with the conceptional differential privacy in common data format, e.g., time-series. We implement our privatizing approach and expose the privatized images to human and machine inspectors to visualize how the privatized images get indistinguishable from each other over privatization setting ranges. The latter inspector quantitatively details the privacy-accuracy trade-off and further reveals that for a given privacy budget, images with more privatized attributes do not necessarily lead to more explicit image semantics, and this impact differs across databases. From both research and application perspectives, such an impact need to be considered to gain a desirable privacy-accuracy trade-off without falling into potential ``pitfall" regions.

\section*{Acknowledgement}

The computations were enabled by resources provided by the Swedish National Infrastructure for Computing (SNIC) at C3SE partially funded by the Swedish Research Council through grant agreement no. 2018-05973.

\bibliographystyle{IEEEtran}
\bibliography{bibfile}

\end{document}